# Linear and Non-Linear Rheology of Single- and Double-Cross-Linked Biopolymer Networks under Viscous Shear Flow

Nasrollah Hajaliakbari[1*], David Head[1], Oliver G Harlen[2]

scnhv@leeds.ac.uk*(corresponding author), d.head@leeds.ac.uk, o.g.harlen@leeds.ac.uk

[1] School of Computer Science, Faculty of Engineering and Physical Science, University of Leeds, Leeds, UK

[2] School of Mathematics, Faculty of Engineering and Physical Science, University of Leeds, Leeds, UK

## Abstract

In this research study, a numerical tool, which is based on a version of Slender Body theory, has been used and also modified to simulate the mechanical behaviour of single- and double-cross-linked biopolymer networks (hydrogel) under oscillatory shear flow. The hydrodynamic interactions among fibres of intertwined networks were considered. Then, the stress and Fourier coefficients (i.e. shear moduli) were evaluated for both linear and nonlinear regimes. It was found that the double peaks (two-step yielding) of two double network at 100% maximum strain amplitude (nonlinear regime) cannot happen due to changes in fibre alignments and seed numbers, although the crosslinkers between two subnetworks present, which was previously reported in the literature. In fact, we also observed two peaks for single network in nonlinear regime.
Furthermore, it was shown that the stress-strain curve of double network is not predicted by just superimposing the results from the corresponding single networks at 5% maximum strain amplitude (linear regime), but this prediction can be provided at 100% maximum strain amplitude (nonlinear regime). The Fourier coefficients and corresponding amplitude (an indication of nonlinearity effects) for double network were quite considerable from zero to fifth modes in nonlinear regime, despite enough zero and first modes in linear regime. It was also shown that the nonlinearity effects can be related to the morphology of the initial structure, i.e. the seed number rather than the flow condition for the single network. These results can help scientists to better design enhance fibrous materials used in wound healing or tissue engineering.

## Introduction

Over the past few decades, significant research efforts have focused on elucidating the mechanical properties of bio-inspired fibrous biomaterials to enable improved design, fabrication, and functional performance. Representative systems include actin networks, collagen-based soft hydrogels, and fibrin networks, which are widely utilized in applications such as wound healing, tissue engineering, sensing, and drug delivery. In fact, the microstructure of such these intricate complex networks governs the whole macromechanical properties, which must be accurately modelled. In 2007, the single biopolymer network such as actin network has been analytically modelled via a modified worm-like chain (WLC) to consider the effects of fibre extensibility on 3D multiaxial stress-strain behaviour (1). It was shown that as the persistence length increases, the shear stress and tangent modulus would rapidly increase. The linear force–extension relationship was not able to accurate capture the strain stiffening behaviours found in experiments.

A theoretical tool was developed by Ban et al. (2) to consider the abnormal changes in Poisson effects on the stiffness of collagen networks when they are stretched triaxially. This model was developed by defining the strain energy density function for a neo-Hookean material to consider the coupling of deformation and the Poisson effect simultaneously when both shear and volumetric changes present in the test. Some finite-element simulations were also performed to capture the effects of anisotropic distributions of displacement in cellular contractions. Some experiments have been conducted to evaluate the viscoelastic stress-strain response and the deformation of fibrin network during unidirectional tensile tests (3). A combination of the computational model with finite element analysis was also used to predict the multiscale mechanical properties of fibrin gel. It was found that different material properties, for example higher concentration of gel would lead to higher modulus and the fibrin gel undergoes non-affine reorientation.

Some experiments have been conducted on both single and double biopolymer network to evaluate he effects on crosslinking on shear-strain curve (4). Two double peaks (two-step yielding) were found in the samples at high strain rates for Alg+/PAAm- hydrogel (the Alginate (Alg) network is crosslinked but not polyacrylamide (PAAm) network) due to hydrogen bonds. In 2025, the mechanical behaviour of two types of double network gels under shear flow have been very recently investigated by Mugnai et al (5). They implemented a numerical method to evaluate the effects of demixing order parameter, average pore size and contacts on linear and nonlinear rheological behaviours. They reported that the interspecies interactions can become more dominant for intertwined double network. In their computer simulations using LAMMPS, intraspecies interactions are kept fixed and only interspecies interactions are varied to sperate distinct species effects. By this assumption, they ignored the effects of intraspecies that can be important. Therefore, we aimed to fill this gap and systematically study three types of single and two types of double intertwined network under oscillatory shear flow. We now consider both the effects of intra fibre interaction and inter fibre interaction via crosslinking to predict the stress for both linear and nonlinear rheology.

## Materials and method

The dynamics of sedimenting fibres in a viscous fluid has been previously studied using the simulation tool developed by Maxian et al.(6-8). They used Rotne-Prager-Yamakawa (RPY) kernels to evaluate the hydrodynamic interactions among Brownian fibres. This tool was also capable of capturing the evolution of an assembly of non-Brownian crosslinked fibres under shear flow for biopolymers like actin (9, 10). In fact, a modified version of Slender Body Theory for Stokes flow was used and modified in a way that it was capable of simulating the mechanical behaviour of both single and double network under various shear flow conditions.

## Fibre approximation (Chebyshev points and tangent vectors)

A type 2 Chebyshev polynomial interpolation can be made among $m_x$ distinct points ($\boldsymbol{X}$) to approximate the fibre profile as a continuous function. Similarly, a type of 1 Chebyshev polynomial interpolation can be made among $m$ distinct tangent vectors, which are represented by $\boldsymbol{\tau}$ and are equal to $m_\text{x} = m + 1$, to approximate a continuous tangent function required in calculations. This discretised set of points $\boldsymbol{X}$ have been shown in Figure 1 and are related to the midpoint and the tangent vectors $\boldsymbol{\tau}$ by the invertible matrix $\boldsymbol{\chi}$

$$\boldsymbol{X} = \boldsymbol{\chi} \begin{pmatrix} \boldsymbol{\tau} \\ \boldsymbol{X}_{MP} \end{pmatrix}. \tag{1}$$

The operator $\boldsymbol{\chi}$ maps tangent vectors and the midpoint into the Chebyshev points, i.e. positions

$$\boldsymbol{X} = \left(\boldsymbol{D}_{\boldsymbol{m+1}}^{\dagger}\boldsymbol{E}_{\boldsymbol{m\to m+1}} \quad \boldsymbol{B}\right)\begin{pmatrix}\boldsymbol{\tau} \\ \boldsymbol{X}_{\boldsymbol{MP}}\end{pmatrix} := \chi\bar{\boldsymbol{\tau}} \tag{2}$$

The matrix $\boldsymbol{B}$ in the first parenthesis is defined such that $\boldsymbol{X}_{\boldsymbol{MP}}$ is the middle point (midpoint) of $\boldsymbol{X}$ on $m_x$ Chebyshev grid number. The second parenthesis includes the tangent vectors plus the midpoint. In the first parenthesis, an extension matrix $\boldsymbol{E}$ must be used to take the tangent vectors from a grid of size $m$ into one of size $m_x$. $\boldsymbol{D}$ is differentiation matrix that must be done on the extended grid size, i.e. $m+1$, that is $m_x$ (shown in Figure 1). The superscript † shows the pseudo-inverse of this matrix is required for mapping. The inverse of operator $\boldsymbol{\chi}$ can be used to differentiate $\boldsymbol{X}$ on the grid size $m_x$ and then downsamples to the grid size of $m$ via the contraction matrix $\boldsymbol{E}_{\boldsymbol{m+1\to m}}$ .

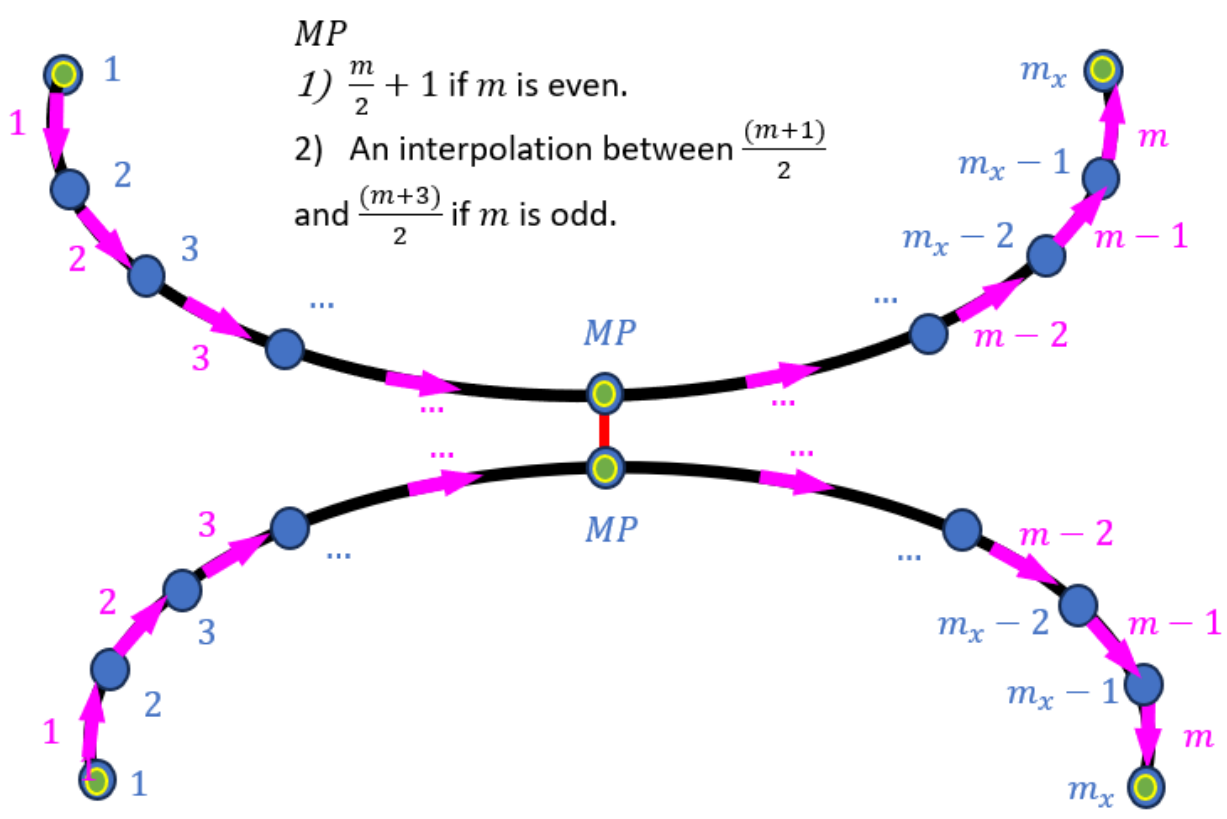

Figure 1. The schematic illustration of discretised fibre described by Chebyshev points $\boldsymbol{X}$ (blue and black solid circles) and tangent vectors $\boldsymbol{\tau}$ (magenta arrows). The solid black line represents the continuous profile of the fibre, $\mathbb{X}$ (Chebyshev interpolant). The uniform points are illustrated by green circles with yellow line and the crosslinker has been shown as a red solid line.

## Crosslinkers (uniform points)

The Chebyshev points are not uniformly distributed along the fibre. Duo to this property, a new set of points ($N_u$), represented as a green circle with yellow line in Figure 1, are required to be uniformly established along the fibre (known as binding sites) to allow that fibre to bind another fibre with a crosslinker (red solid line in Figure 1). This process would lead to construction of a percolated network with multiple fibres. Each crosslinker is modelled as a spring between a pair of uniform points that belongs to a pair of separate fibres. This is done by resampling the fibre at $N_u$ uniform points to form the vector $\overline{\boldsymbol{X}}$ with the use of Chebyshev points $\boldsymbol{X}$ (9),

$$\overline{\boldsymbol{X}} = \boldsymbol{R}^{(u)}\,\boldsymbol{X} \quad \leftrightarrow \overline{\boldsymbol{X}}_{\{p\}} = \boldsymbol{R}_{\{p,:\}}^{(u)}\,\boldsymbol{X} \tag{3}$$

Where $\boldsymbol{R}^{(\boldsymbol{u})}$ is the resampling matrix and $\boldsymbol{R}_{\{p,:\}}^{(u)}$ is its $p$th row.

The displacement between two uniform points ($p$th uniform point belongs to fibre $i$ and $q$th uniform point that is located on fibre $j$) is evaluated by

$$\boldsymbol{r} = \overline{\boldsymbol{X}}_{\{p\}}^{(i)} - \overline{\boldsymbol{X}}_{\{q\}}^{(j)} = \boldsymbol{R}_{\{p,:\}}^{(u)}\,\boldsymbol{X}^{(i)} - \boldsymbol{R}_{\{q,:\}}^{(u)}\,\boldsymbol{X}^{(j)} \tag{4}$$

Where $r = \|\boldsymbol{r}\|$ is defined as the length of that displacement. The energy that is stored by a crosslinker is calculated by,

$$\varepsilon(r) = \frac{K_c}{2}(r-\ell)^2 \tag{5}$$

The energy gradient is needed to be calculated before determining the crosslinker force,

$$\frac{\partial \varepsilon(r)}{\partial r} = K_c(r - \ell) \tag{6}$$

Note that $K_c$ and $\ell$ are the constant stiffness and rest length of the crosslinker spring. The crosslinker forces between a pair of uniform points on fibre i and j are obtained by,

$$\begin{aligned} \boldsymbol{F}_{\{a\}}^{(i)} &= \frac{\partial \varepsilon}{\partial r}\frac{\partial r}{\partial \boldsymbol{X}_{\{a\}}^{(i)}} = -\frac{\partial \varepsilon}{\partial r}\hat{\boldsymbol{r}}\boldsymbol{R}_{\{p,a\}}^{(u)} \\ \boldsymbol{F}_{\{b\}}^{(j)} &= \frac{\partial \varepsilon}{\partial r}\frac{\partial r}{\partial \boldsymbol{X}_{\{b\}}^{(j)}} = -\frac{\partial \varepsilon}{\partial r}\hat{\boldsymbol{r}}\boldsymbol{R}_{\{q,b\}}^{(u)} \end{aligned} \tag{7}$$

A Dynamic cross-linking model was initially used to create a percolated structure using the Brownian motion. There are four reactions which are considered in this model: first end biding, second end binding, first end unbinding and second end binding. For this mode, four binding and unbinding rates must be set to be nonzero. Once the structure was created, all these rates are set to be equal to zero and the resultant structure is put under shear flow. This setting leads to not binding or unbinding events within the structure, which is important to keep the connectivity of the structure when shear strain is imposed (permanent links). Further details about reactions and crosslinking mechanism have been provide in (10, 11).

## Fibre dynamics

As abovementioned, a set of Chebyshev points ($\boldsymbol{X}$) can be used to discretise the fibre profile. Then, an interpolant is mathematically obtained as a continuous function, $\mathbb{X}(s,t)$, with $s$ the distance along the contour, $0 \leq s \leq l$.

$$\mathbb{X}(s,t) = \boldsymbol{X}_{MP}(t) + \int_{\frac{l}{2}}^{s} \mathbb{T}(s',t)\,ds'. \tag{8}$$

Where $\mathbb{T}(s) = \frac{\partial}{\partial s}\mathbb{X}$ is tangent vector and $\boldsymbol{X}_{MP}$ is the position of fibre midpoint (the discretised distribution of $\mathbb{X}$, evaluated at the midpoint $MP$, $s = l/2$). Parameter $l$ is the length of the fibre. As the equation (8) is a continuous function, it can be differentiated (with respect to the time $t$) and converted into discretised form, i.e.

$$\partial_t \boldsymbol{X}(s,t) = \boldsymbol{U_{MP}}(t) + \int_{\frac{l}{2}}^{s} \partial_t \boldsymbol{\tau}(s',t)\,ds'. \tag{9}$$

Where $\boldsymbol{U_{MP}}$ and $\boldsymbol{\tau}$ are the fibre midpoint velocity and discretised form of tangent vectors, respectively. It is assumed that the fibre is inextensible. Based on that, the tangent vectors must only be able to rotate, i.e.

$$\partial_t \boldsymbol{\tau} = \boldsymbol{\Omega} \times \boldsymbol{\tau}, \tag{10}$$

Where $\boldsymbol{\Omega}$ is the set of local tangent rotations.

## Mobility equation

The mobility equation has been commonly used in literature to describe the motion of solid body in Stokes flows. Based on that, the fibre motion is linearly related to the sum of force densities acting on it by a mobility operator $\mathbb{M}[\mathbb{X}]$ (which depends on the current configuration of the fibre).

$$\partial_t \mathbb{X}(s,t) - u_0(s) = \mathbb{M}[\mathbb{X}]\mathbb{f}. \tag{11}$$

where $\mathbb{f}$ is the sum of force densities, i.e. the bending, tension, crosslinking force densities.

By considering the inextensibility condition, a saddle point system is formed that is

$$\begin{pmatrix} -\mathbb{M}[\mathbb{X}] & \mathbb{K} \\ \mathbb{K}^* & 0 \end{pmatrix}\begin{pmatrix} \boldsymbol{\lambda} \\ \boldsymbol{\alpha} \end{pmatrix} = \begin{pmatrix} \mathbb{M}[\mathbb{X}]\left(\mathbb{f}^k + \mathbb{f}^{CL} + \mathbb{f}^{St}\right) + u_0[\mathbb{X}] \\ 0 \end{pmatrix} \quad (12)$$

Where $\mathbb{f}^{CL}$ , $\mathbb{f}^{k}$, $\mathbb{f}^{St}$, $\mathbb{K}$ , $\boldsymbol{\lambda}$ and are crosslinker, bending and steric force densities and kinematic operator, respectively.

The mobility operator can be obtained by use of some singularities described by a modified version of Slender Body Theory. Based on that, the fibre velocity in equation (11) can be evaluated by

$$\boldsymbol{U}_L(s) = u_0\left(\mathbb{X}(s)\right) + \frac{1}{8\pi\mu}\Bigg( \left(c(s)\left(\boldsymbol{I} + \boldsymbol{\tau}(s)\boldsymbol{\tau}^T(s)\right)\right) + \left(\boldsymbol{I} - 3\boldsymbol{\tau}(s)\boldsymbol{\tau}^T(s)\right)\right)\mathbb{f}(s) + \\ \int_0^l \left(\left(\frac{\boldsymbol{I} + \hat{\boldsymbol{r}}(s,s')\hat{\boldsymbol{r}}^T(s,s')}{r(s,s')}\right)\mathbb{f}(s') - \left(\frac{\boldsymbol{I} + \boldsymbol{\tau}(s)\boldsymbol{\tau}^T(s)}{|s-s'|}\right)\mathbb{f}(s)\right) ds' \Bigg), \quad (13)$$

Where $\boldsymbol{r}(s,s') = \boldsymbol{X}(s) - \boldsymbol{X}(s')$, $r = \|\boldsymbol{r}\|$, and $\hat{\boldsymbol{r}} = \boldsymbol{r}/r$. The velocity includes two terms: local drag (first line) and finite part (the second line), i.e. the intra-fibre nonlocal hydrodynamic interactions. The local drag coefficient $c(s)$ is defined as

$$c(s) = \ln\left(\frac{4(l-s)s}{a^2}\right), \quad (14)$$

## Bending force density

The bending force density is proportional to the fourth derivative of the continuous fibre profile with respect to arc length $s$, i.e.

$$\mathbb{f}^k = -k\partial_s^4\mathbb{X}, \quad (15)$$

where $k$ is the bending stiffness of the fibre. For the single network, which has been already implemented by Maxian et al. (9), the bending stiffness of the fibre is a constant whereas for double network, two different values must be considered. Here, the source code was implemented in a way to consider these effects. Therefore, the bending force density must be defined as,

$$\mathbb{f}_1^k = -k_1\partial_s^4\mathbb{X} \quad (16)$$

$$\mathbb{f}_2^k = -k_2\partial_s^4\mathbb{X}, \quad (17)$$

Where $\mathbb{f}_1^k$, $\mathbb{f}_2^k$, $k_1$ and $k_2$ are the bending force density for network 1 and network 2, fibre bending stiffness for network 1 and network 2 in double network, respectively.

## Steric forces

In the simulations, depending on the setting of parameters, it is quite plausible that some fibres become very close to each other and even cross each other which is nonphysical. To avoid this problem, the steric forces (interactions) must be considered. Similar to crosslinker modelling, the steric interaction energy function between fibres $i$ and $j$ can be defined as

$$\mathcal{E}^{(ij)} = \int_0^l \int_0^l \hat{\varepsilon}\left(r\left(s^{(i)}, s^{(j)}\right)\right) ds^{(i)} ds^{(j)} \tag{18}$$

$$r\left(s^{(i)}, s^{(j)}\right) = \left\|\mathbb{X}^{(i)}\left(s^{(i)}\right) - \mathbb{X}^{(j)}\left(s^{(j)}\right)\right\| \tag{19}$$

Where $\hat{\varepsilon}$ is the potential density function defined as,

$$\hat{\varepsilon}(r) = \frac{\varepsilon_0}{a^2} \operatorname{erf}\left({}^{r}/_{(\delta\sqrt{2})}\right) \tag{20}$$

$$\frac{d\hat{\varepsilon}}{dr} = \frac{\varepsilon_0}{a^2\delta}\sqrt{\frac{2}{\pi}} \exp\left({}^{-r^2}/_{(2\delta^2)}\right) \tag{21}$$

Where both $\delta$ and $\varepsilon_0$ control the decay and intensity of steric forces, respectively.

A new grid is required to be defined by the use of the Chebyshev points that is done through upsampling denoted by $\boldsymbol{X}^{(u)} = \boldsymbol{EX}$. Based on these, the steric forces can be evaluated as

$$\mathcal{E} = \sum_k \sum_j \hat{\varepsilon}\left(\left\|\boldsymbol{X}_{\{k\}}^{(u)} - \boldsymbol{X}_{\{j\}}^{(u)}\right\|\right) w_k w_j = \sum_k \sum_j \hat{\varepsilon}\left(\left\|\boldsymbol{E}_{kp}\boldsymbol{X}_{\{p\}} - \boldsymbol{E}_{jq}\boldsymbol{X}_{\{q\}}\right\|\right) w_k w_j \tag{22}$$

$$\boldsymbol{F}_{\{a\}}^{(i)} = -\frac{\partial \mathcal{E}}{\partial \boldsymbol{X}_{\{a\}}^{(i)}} = \sum_k \sum_j \frac{\partial \hat{\varepsilon}}{\partial r}(r_{kj}) \hat{\boldsymbol{r}}_{kj} \boldsymbol{E}_{ka} w_k w_j \tag{23}$$

Where $F_{\{a\}}^{(i)}$, $w_k$ and $w_j$ are the steric force at Chebyshev node $a$ on fibre $i$, the integration weights of points $k$ and $j$ on the upsampled grid. To do upsampling, a computationally efficient segment-based algorithm was selected. Further elaborations have been given in (9).

## Shear direction

There was a limitation in the research work done by Maxian et al. (9). In fact, there was only one option available to impose the shear flow, that is where the x is the flow direction, y is the gradient direction and z is vorticity direction. Based on that implementation, the transformation between unsheared coordinates $(x, y, z)$ and sheared coordinates $(x', y', z')$ is described as:

$$\begin{pmatrix} x' \\ y' \\ z' \end{pmatrix} = \begin{pmatrix} 1 & -g(t) & 0 \\ 0 & 1 & 0 \\ 0 & 0 & 1 \end{pmatrix} \begin{pmatrix} x \\ y \\ z \end{pmatrix} \tag{24}$$

Where $g(t)$ is the total nondimensional strain. This was not allowed to change the flow condition (direction) and check how isotropic the fibrous network under shear flow is. To alleviate this, the source code is developed in way that there are two other available options in addition to the main case; one where the y is the flow direction, z is the gradient direction and x is the vorticity direction. For this case,

$$\begin{pmatrix} x' \\ y' \\ z' \end{pmatrix} = \begin{pmatrix} 1 & 0 & 0 \\ 0 & 1 & -g(t) \\ 0 & 0 & 1 \end{pmatrix} \begin{pmatrix} x \\ y \\ z \end{pmatrix} \tag{25}$$

Another option is where the z is flow direction, x is the gradient direction and y is the vorticity direction. For this case,

$$\begin{pmatrix} x' \\ y' \\ z' \end{pmatrix} = \begin{pmatrix} 1 & 0 & 0 \\ 0 & 1 & 0 \\ -g(t) & 0 & 1 \end{pmatrix} \begin{pmatrix} x \\ y \\ z \end{pmatrix} \tag{26}$$

With all three options, it is now possible to identify how isotropic the fibrous network under shear flow would behave.

## Results

Three types of single and two types of double networks ($\frac{\chi_1}{\chi_2} = \frac{10}{2}$ *and* $\frac{10}{5}$), depending on non-dimensional bending stiffness of fibres for each subnetwork ($\chi_1$ and $\chi_2$) have been considered here. The initial configuration (structure), that was used to put under oscillatory shear flow, was primarily constructed with fibres with $\chi = 0.0122$ when only Brownian motion was activated. This happened for three seed numbers to construct a network with various morphologies. This process was selected since there was quite extensive range of parameters and it was almost impossible to change all parameters at the same time and see the effects of them on shear metrics. The set of physical parameters has been given in Table 1.

Table 1. The set of physical and numerical parameters used in the simulations

| Parameter | Symbol | Parameter | Symbol |
|---|---|---|---|
| Fibre length | $(l)$ | CL spring stiffness | $(K_c)$ |
| Fibre radius | $(a)$ | CL rest length | $(\ell)$ |
| Fibre persistence length | $(l_p)$ | CL first end binding rate | $(k_{\text{on}})$ |
| Final simulation time | $(T_f)$ | CL second end binding rate | $(k_{\text{off}})$ |
| Length of domain | $(L_d)$ | CL first end unbinding rate | $(k_{\text{on,s}})$ |
| The thermal energy | $(kT)$ | CL second end unbinding rate | $(k_{\text{off,s}})$ |
| Fluid dynamic viscosity | $(\mu)$ | Shear flow frequency | $(f)$ |
| | | Maximum strain amplitude | $(\gamma_0)$ |

These parameters can be nondimensionalised (6, 8, 12), provided in Table 2.

Table 2. The set of nondimensionalized parameters used in the simulations

| Brownian cases | | Shear simulations (non-Brownian cases) | |
|---|---|---|---|
| Parameter | Value | Parameter | Value |
| $\kappa = \frac{l_p}{l}$ | 20, 10 and 4 | $K_c^* = \frac{K_c}{\mu f l}$ | 800 |
| $\tau^* = \frac{l^3\mu}{T_f kT}$ | 0.4065 | $\chi = \frac{2a\mu\gamma_0 l^3}{EI}$ | 0.00244, 0.00488, 0.0122 |
| $k_{on}^* = f\ell k_{\text{on}}, k_{off}^* = f\ell k_{off}$ | 50,0.1 | $k_{on}^* = f\ell k_{\text{on}}, k_{off}^* = f\ell k_{off}$ | 0,0 |
| $k_{on,s}^* = f\ell k_{\text{on,s}}, k_{off,s}^* = f\ell k_{\text{on,s}}$ | 50,0.1 | $k_{on,s}^* = f\ell k_{\text{on,s}}, k_{off,s}^* = f\ell k_{\text{on,s}}$ | 0,0 |
| Common parameter | | $\ell^* = \frac{\ell}{l}$ | 0.02 |
| Parameter | Value | $\gamma_0^* = \frac{\gamma_0}{f}$ | 1, 0.05 |
| $\epsilon = \frac{a}{l}$ | 0.008 | $T_f^* = T_f f$ | 20 |
| $l_d^* = \frac{L_d}{l}$ | 1.5 | | |

The values of discretisation parameters have been given in Table 3.

Table 3. The set of discretisation parameters

| Parameter | Value | Parameter | Value |
|---|---|---|---|
| Number of Chebyshev points $(N)$ | 9 | Number of shear cycles $(n_{cyc})$ | 20 |
| Number of binding points (uniform points) $(N_s)$ | 3 | Time step size $(\Delta t)$ | $10^{-5}$ |

The simulation was continued and the average of maximum position of total fibres was monitored until it reached a steady value (steady state condition) and percolation was also satisfied in three directions (13). The final configuration constructed by Brownian motion has been plotted in Figure 2.

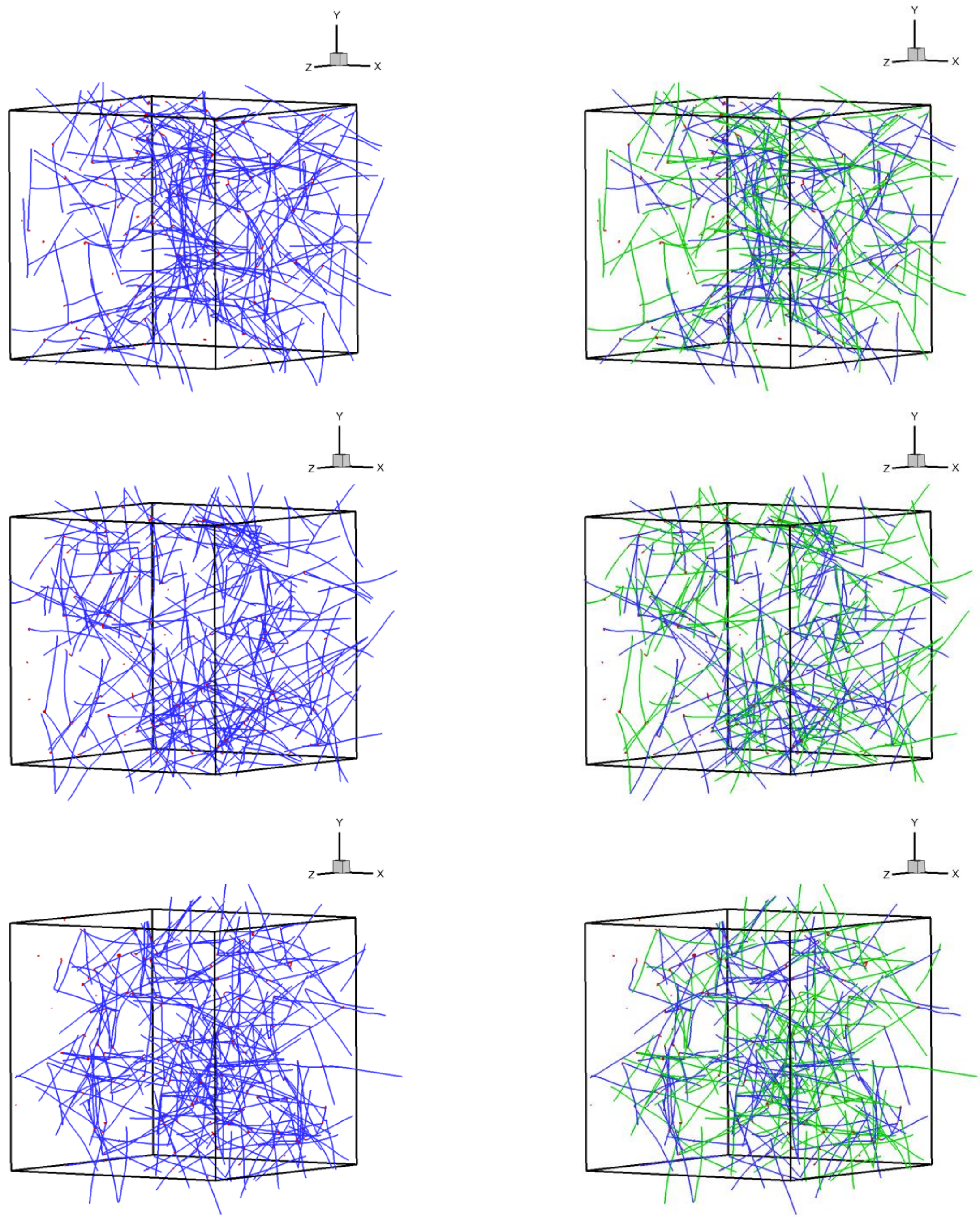

Figure 2. The constructed single (left images) and double (right images) networks for three seed numbers

The constructed networks were used in all shear simulations including the double network hydrogel.

## Stress-strain relation

Three values of persistence length were considered to mimic the mechanical behaviour of the soft ($l_p = 2$), medium ($l_p = 5$) and stiff ($l_p = 10$) fibres under shear flow. The simulations were done for 20 cycles at frequency of 1 Hz. It was observed that this number of cycles are enough to eliminate the transient effects (after 10 cycles, the transient effects were completely disappeared) of the deformed structure under shear flow. The Brownian motion was deactivated and the intra-fibre hydrodynamic for the fibres was only considered.

However, due to presence of crosslinkers (connectivities), it can be assumed that the nonlocal hydrodynamics was kind of considered. The total stress against strain in the last cycle of the network with the stiff fibres, i.e. $l_p = 10$ has been depicted in Figure 3.

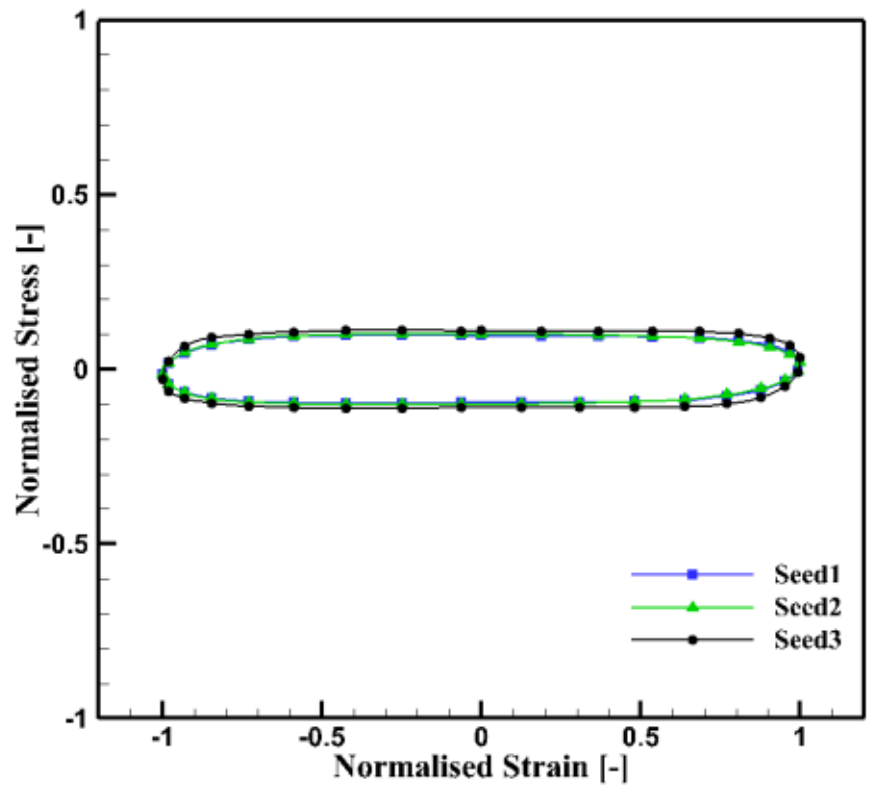

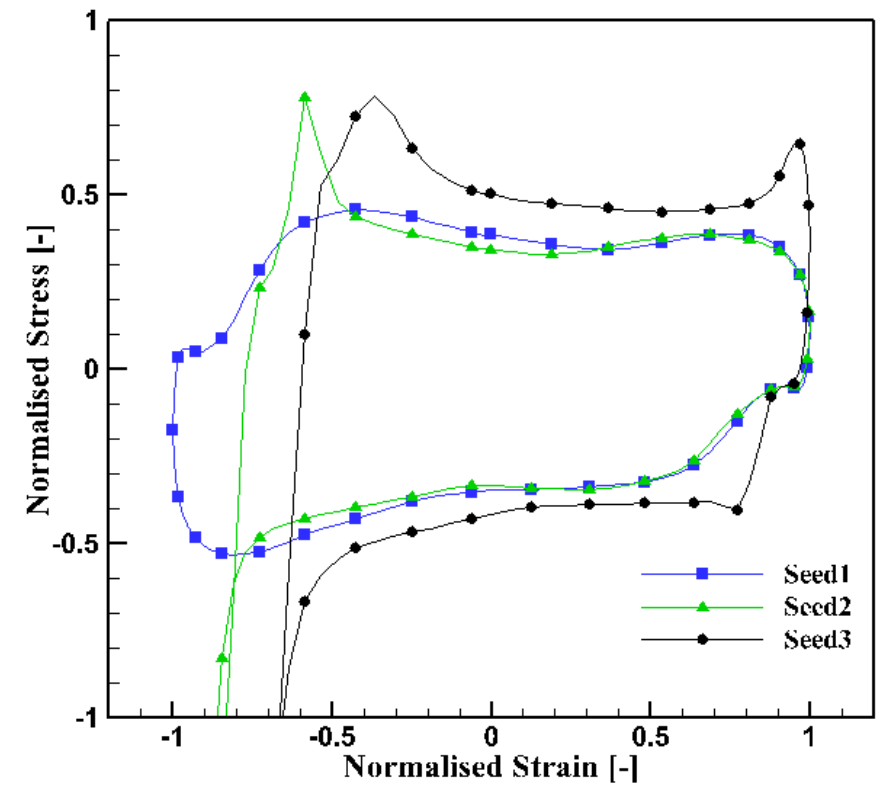

Figure 3. The normalized stress plotted against normalized strain for three networks (three seed numbers) built with the stiffest fibres ($l_p = 10$) at 5% (left image) and 100% (right image) maximum strain amplitude

The stress and strain were normalised by maximum strain amplitude (5% or 100%) for each case. As shown in the left image in Figure 3, in linear regime, the stress-strain follows a horizontal elliptic shape as the maximum strain rate is low (5 percent). This shape remains almost constant even for various seed numbers. However, it is not the same for various seed numbers when the maximum strain rate increases (100 percent), demonstrated in right image of Figure 3. The stress-strain curve shows double peaks at high strain rate, i.e. Figure 3 (right image). The exact constructed network was put under different shear direction to prove that the constructed networks are quite isotropic. The results for the single network with seed 1 constructed with the stiffest fibres have been illustrated in Figure 4.

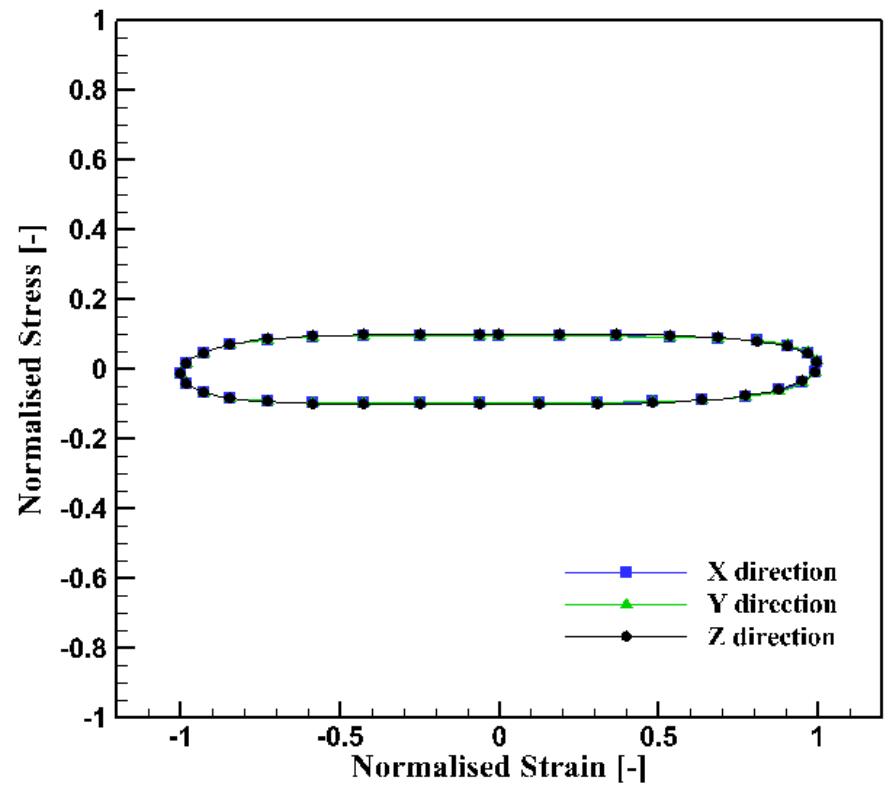

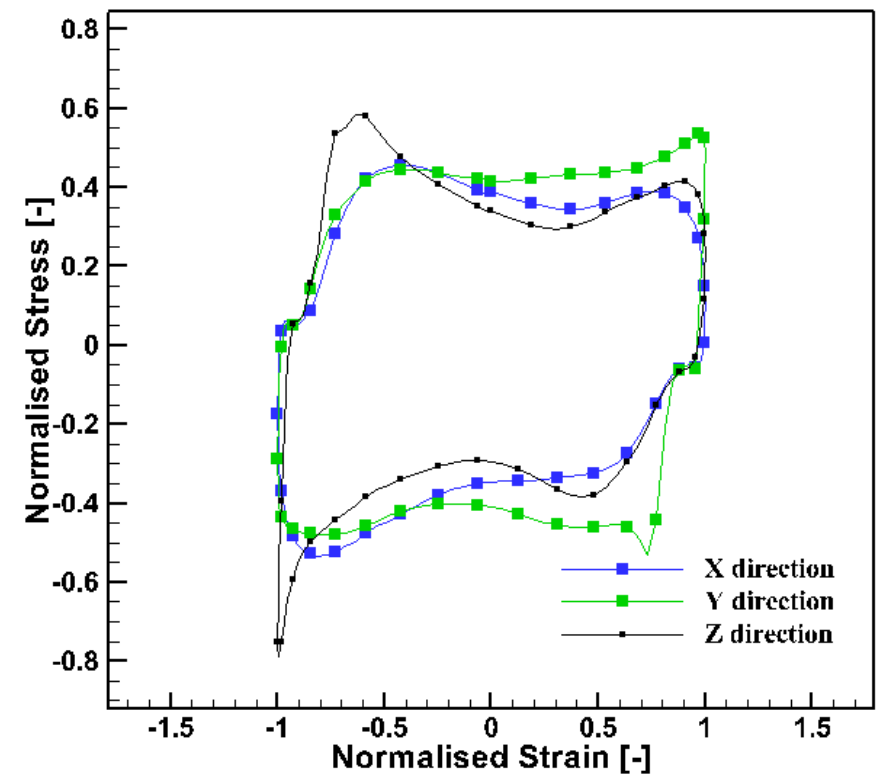

Figure 4. The normalized stress plotted against normalized strain for the single network with seed 1 built with the stiff fibres ($l_p = 10$) at 5% (left image) and 100% (right image) maximum strain amplitude when the flow direction changes

In general, the results are quite similar for both linear and nonlinear regimes as the flow direction changes. This shows that the constructed network is isotropic even when the flow condition (direction) changes.

Two double networks were simulated in this research study: One for which, the persistence length of 10 for 150 fibres and 5 for the remaining (150 fibres) were considered, for another one, instead of setting the persistence length of 2 for soft fibres, 5 was replaced. The total stress against strain has been plotted for these two types of double network hydrogel. Again, the Brownian motion was deactivated and the local hydrodynamic for the fibres was only considered. The total stress was plotted against the strain in Figure 5.

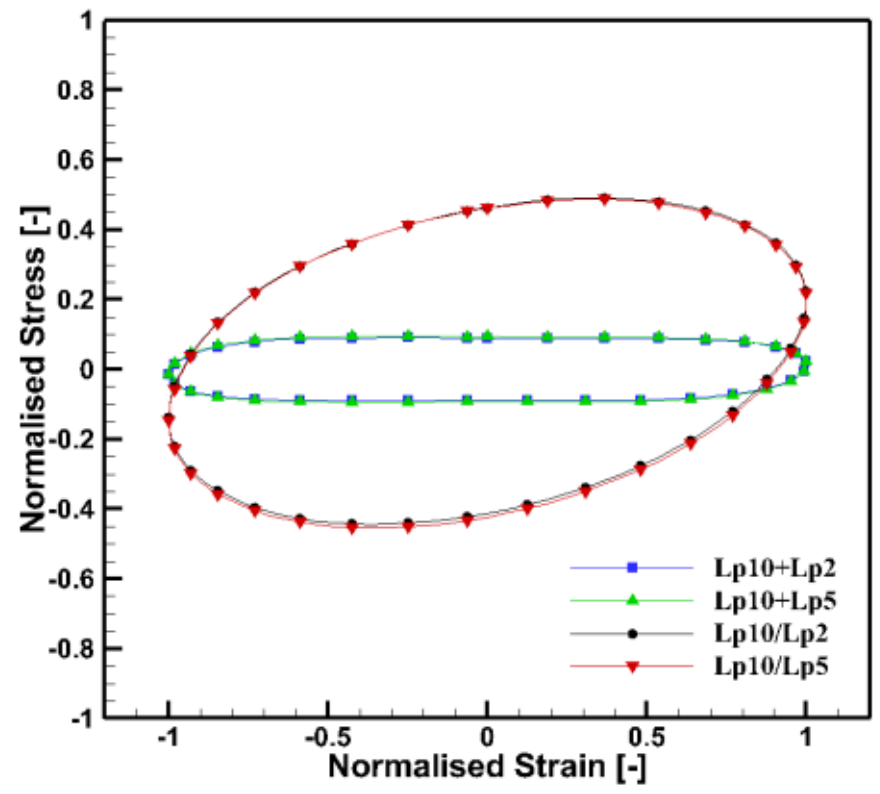

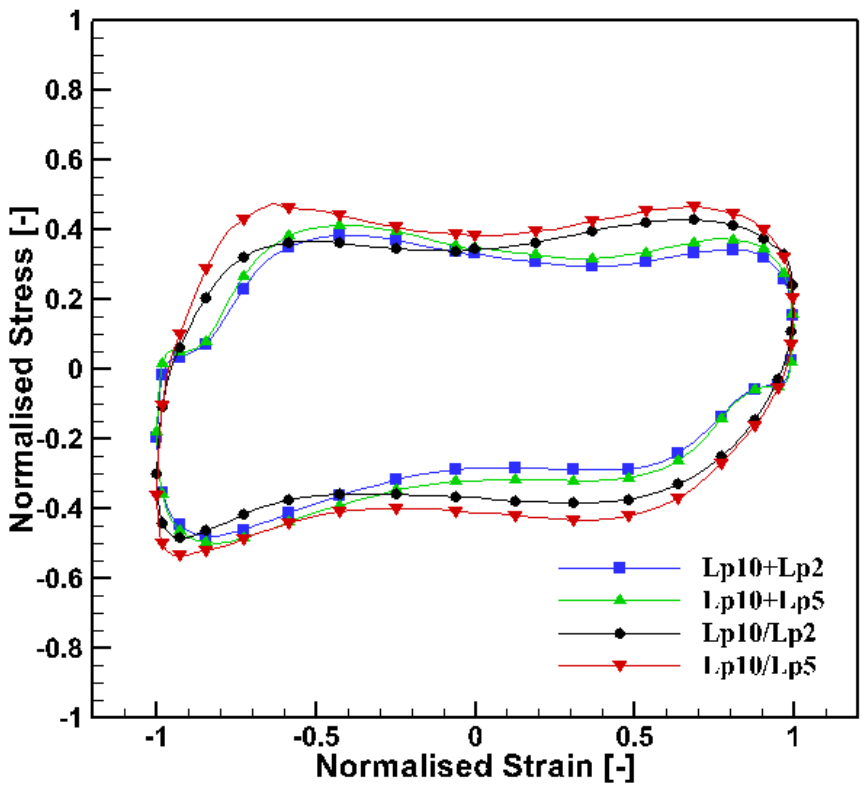

Figure 5. The normalized stress plotted against normalized strain for two double networks ($l_p10/l_p2$ and $l_p10/l_p5$ ) and two double-hypothesized networks ($l_p10 + l_p2$ and $l_p10 + l_p5$ ), constructed from the relevant single networks at 5% (left image) and 100% (right image) maximum strain amplitude for seed 1 when $x$ is the flow direction

There are two double-hypothesized networks in the results shown here ( $l_p10 + l_p2$ and $l_p10 + l_p5$ ), which are provided from results obtained from corresponding single networks. In fact, they are created by halving the stress for each belonging single network with related fibre persistence length and adding them together to create each double-hypothesised network. As can be clearly seen, there is a big difference in normalised stress value compared to two double networks when the maximum strain amplitude is small (left image). As a result, the mechanical behaviour of double network is not predicted by superimposing the results from the corresponding single networks. However, this difference diminishes when the maximum strain amplitude is large (right image), i.e. nonlinear regime. The double peaks and double minimums can be clearly identified in nonlinear regime. The double peaks have already been reported by Kopnar et al (4). The double peak was related to the presence of hydrogen bonds (crosslinkers between two networks) for some samples in which the permanent crosslinkers are not present, i.e. the PAAm fibres in the mixture of i.e. Alg+/PAAm- hydrogel are not crosslinked.

However, the binding and unbinding rates were off in the model here. These double peaks can be related to some fibre alignments or seed number of the sample, not just the crosslinkers. As it is observed, there are also double minimums at 100% maximum strain amplitude when the strain direction is reversed. To further evaluate this phenomenon, other scenarios were simulated and stress was decomposed into two elements, each one related to each subnetwork within a double network. The results have been shown in Figure 7.

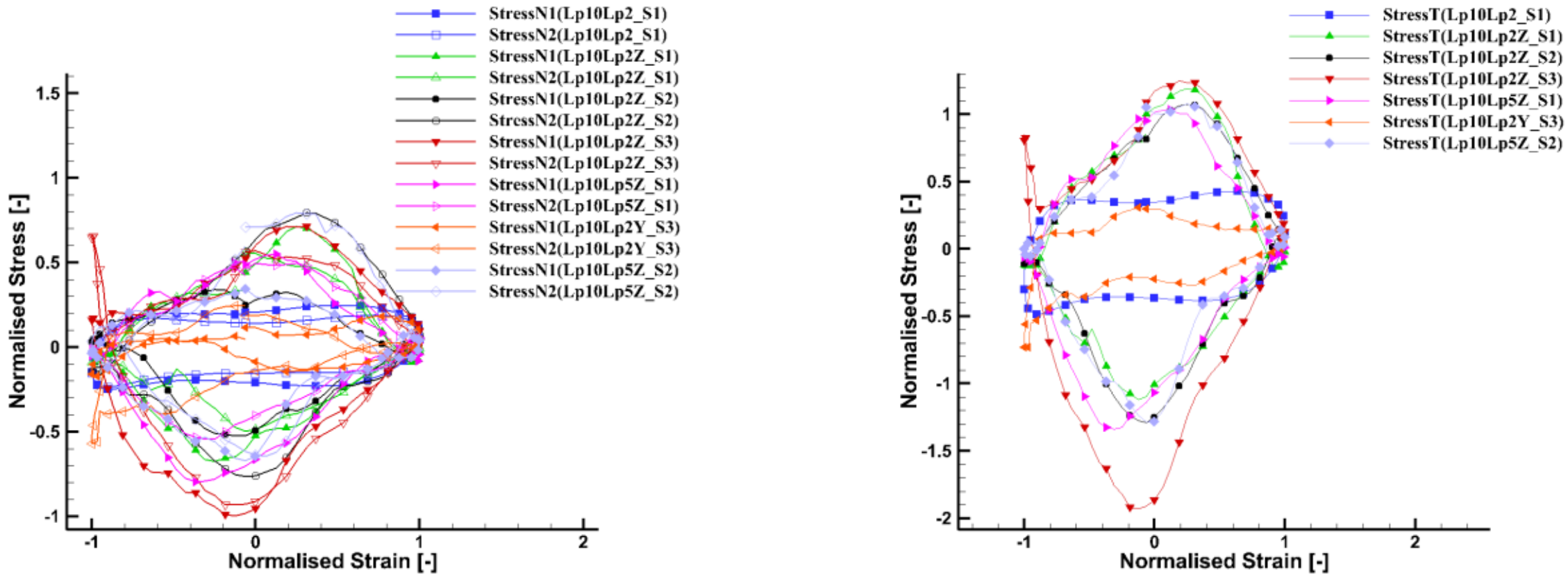

Figure 6. The decomposed (left image) and total (right image) normalized stress plotted against normalized strain for two double networks ($l_p10/l_p2$ and $l_p10/l_p5$ ) under various flow conditions and seed numbers at 100% (right image) maximum strain amplitude

The stress decomposition for double network with persistence fibre ratio $l_p10/l_p2$ at seed 1 and when the flow direction is x, clearly shows double peaks for each subnetwork (the filled and hollow square with blue line in the left image), as well as the total stress (filled square with blue line in the right image). However, there is no double peak in total stress (green line with delta symbol in the right image) for the exact double network when the flow direction changes from $x$ to $z$ or seed number changes from 1 to 2 (black line with circle in the right image). In the right image, the green line with the filled delta shows one peak instead of double peak, whereas for its decompositions in the right image, only network 2 with soft fibres has a double peak (the green line with hollow delta), despite one peak for the network 1 with the stiff fibres (the green line with filled delta).

## Fourier coefficients and amplitude

To evaluate how nonlinear the response of a structure to impose shear is, the Fourier coefficients were calculated for the ten modes of the last five cycles. The coefficients $a_i$ and $b_i$ are related to the $Cos$ and $Sin$ functions, respectively, while the imposed strain is just a function of $Sin$. These Fourier coefficients and the amplitude ($\sqrt{a_i^2 + b_i^2}$) for single network built with the stiff fibres were plotted in Figure 7.

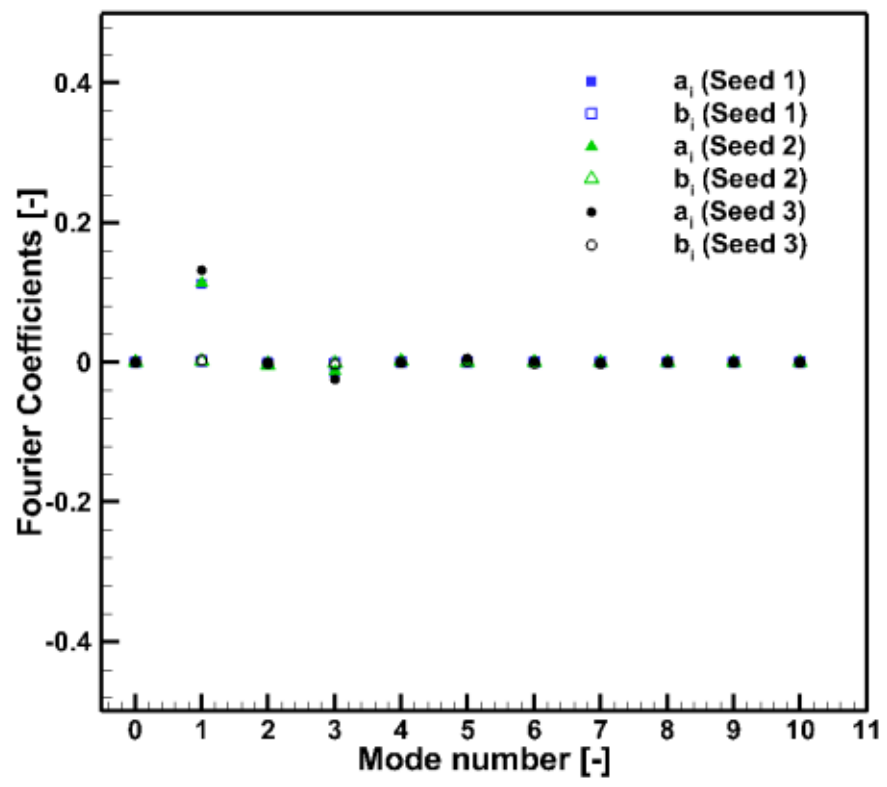

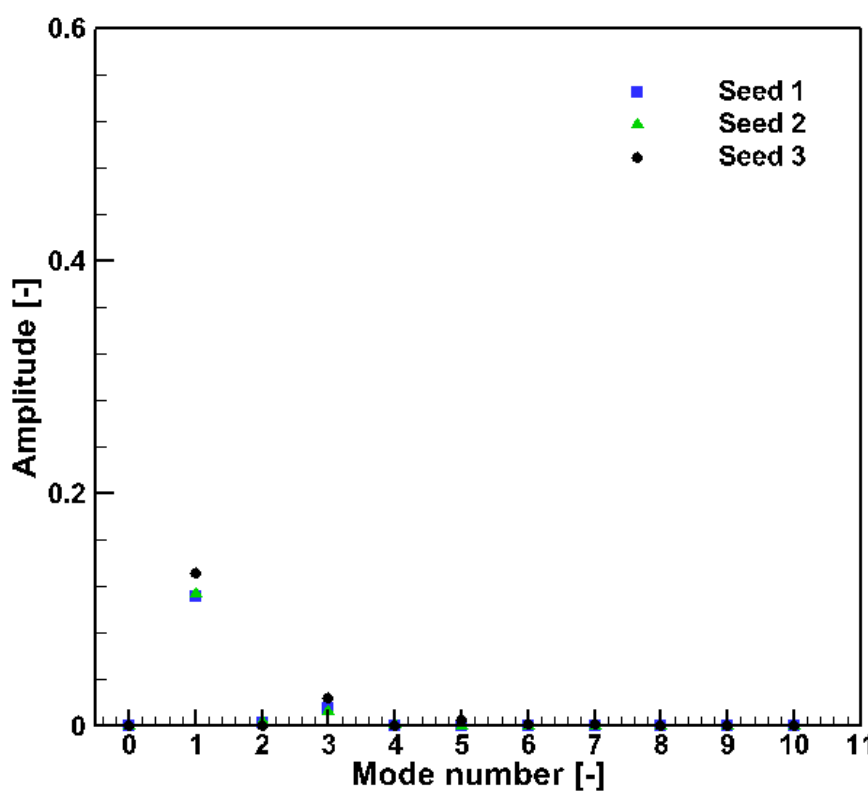

Figure 7. The Fourier coefficients (left image) and amplitude (right image) obtained from normalized stress plotted against mode number for three networks (three seed numbers) built with the stiff fibres ($l_p = 10$) at 5% maximum strain amplitude

The $a_i$ values of the first mode are the only large ones and others can be ignored (left image). The amplitude (right image) is comparable for the first and somehow third modes. These show that the response is almost quite linear at 5% maximum strain amplitude even when the seed number changes. These Fourier coefficients and amplitude have been demonstrated in Figure 8 at 100% maximum strain amplitude.

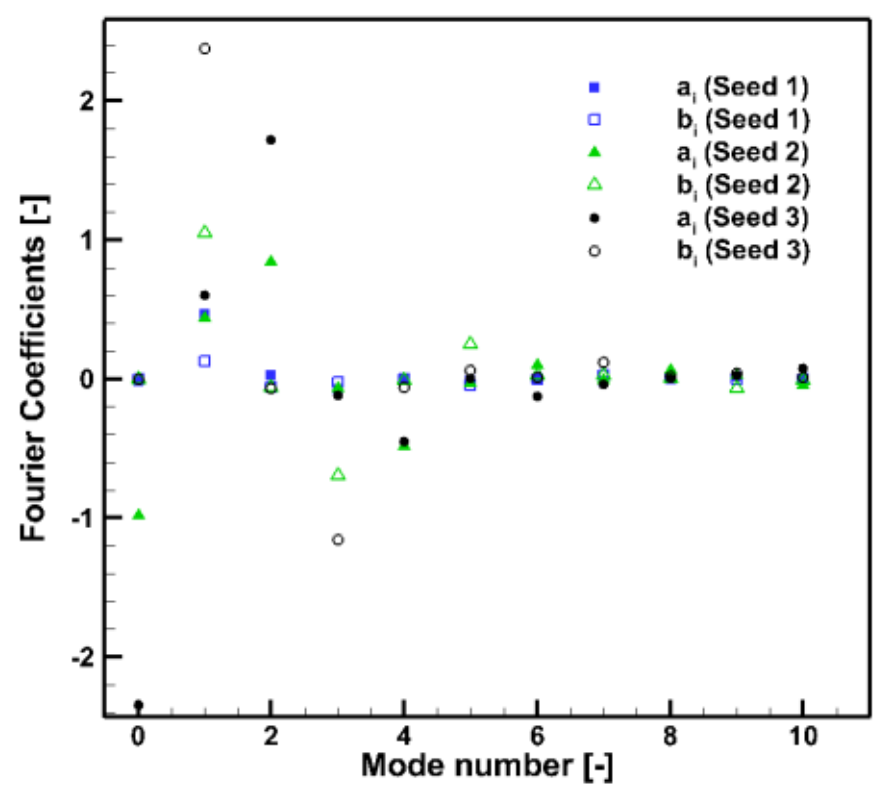

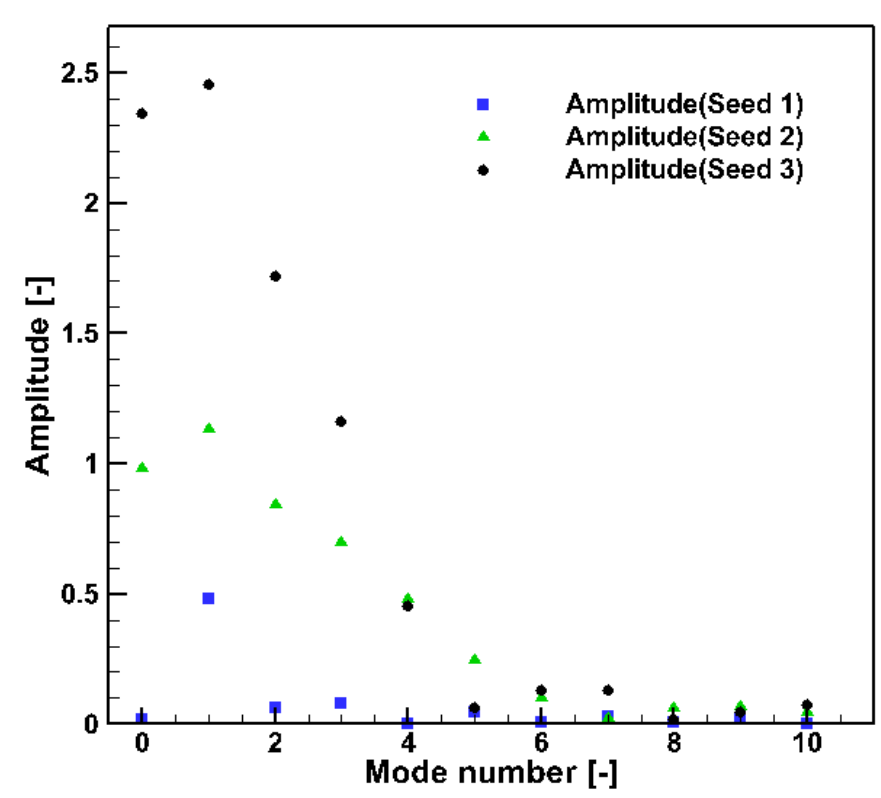

Figure 8. The Fourier coefficients (left image) and amplitude (right image) obtained from normalized stress plotted against mode number for three networks (three seed numbers) built with the stiff fibres ($l_p = 10$) at 100% maximum strain amplitude

As it is observed in left image, the Fourier coefficients are considerable from the zero until the fifth modes when the maximum strain amplitude is 100%. They are also quite scattered as the seed number changes. The amplitude is quite large for the first mode, but it starts to decreases as the mode number increases. The difference in the amplitude (right image) of the initial modes is also large as the seed number changes. These differences become less pronounced when the flow direction changes (Figure 9).

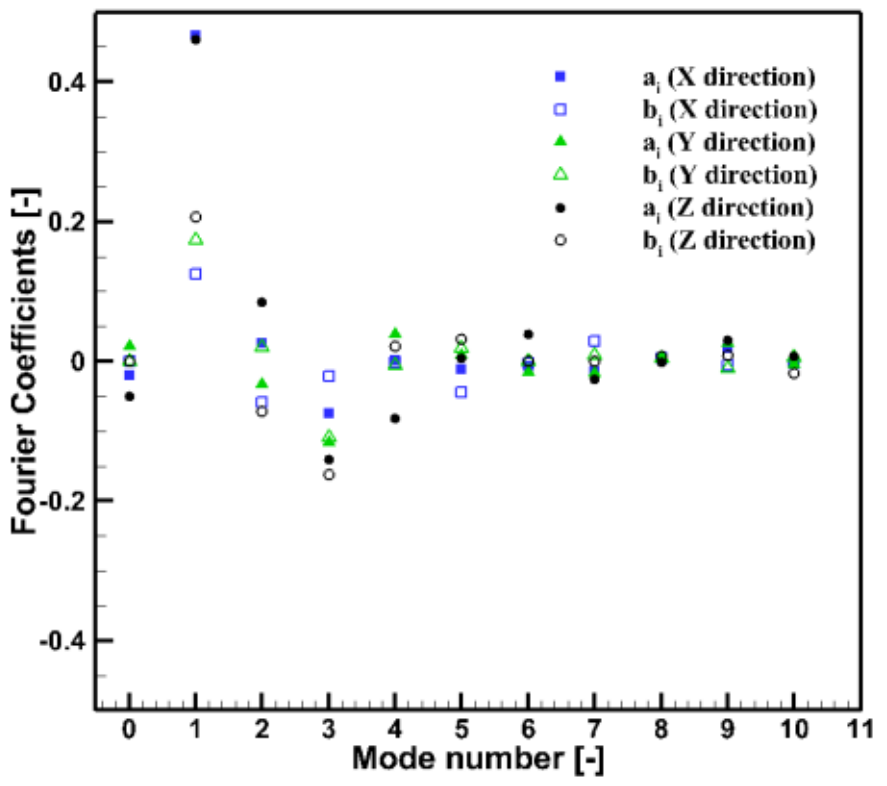

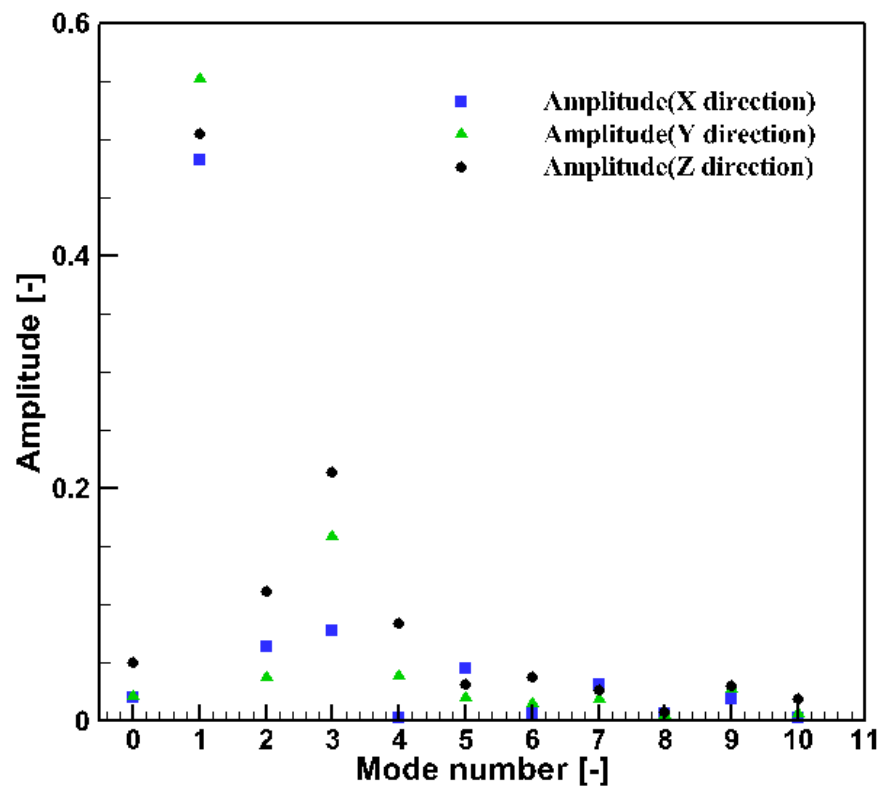

Figure 9. The Fourier coefficients (left image) and amplitude (right image) obtained from normalized stress plotted against mode number for the single network with seed 1 built with the stiff fibres ($l_p = 10$) at 100% maximum strain amplitude for when the flow direction changes

It is clear that the nonlinearity effects can be related to the morphology of the initial structure, i.e. the seed number rather than the flow condition.
The Fourier coefficients and amplitude for double networks have been plotted in Figure 11 and Figure 11 for 5% and 100% maximum strain amplitudes, respectively.

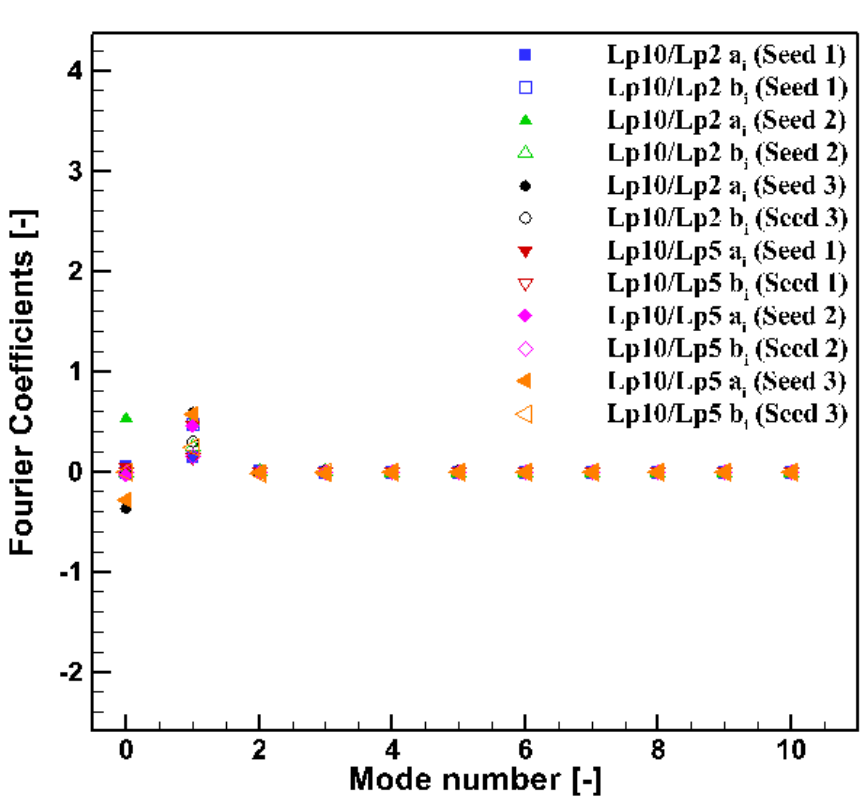

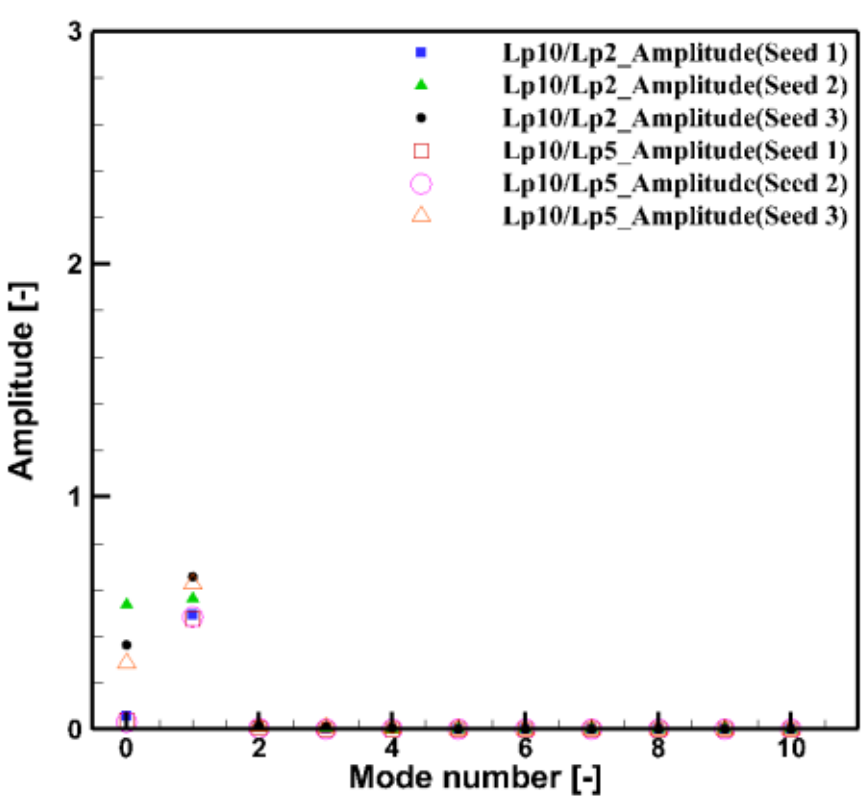

Figure 10. The Fourier coefficients (left image) and amplitude (right image) obtained from normalized stress plotted against mode number for two double networks with three seed numbers at 5% maximum strain amplitude

Compared to single network with the stiff fibres (the right image in Figure 7), the amplitude does not change so much when the seed number changes for the first mode. However, for some seed numbers, the zero mode is large. This can also be seen for Fourier coefficients in the zero mode of two double networks (the left image in Figure 10) compared to single network (the left image in Figure 7).
When the maximum strain amplitude is 100%, the Fourier coefficients are needed to be considered from zero until fifth modes (the left image of Figure 11), similar to single network (the left image of Figure 8). The amplitude does not change some much between two double networks to at the same seed number (the right image of Figure 11).

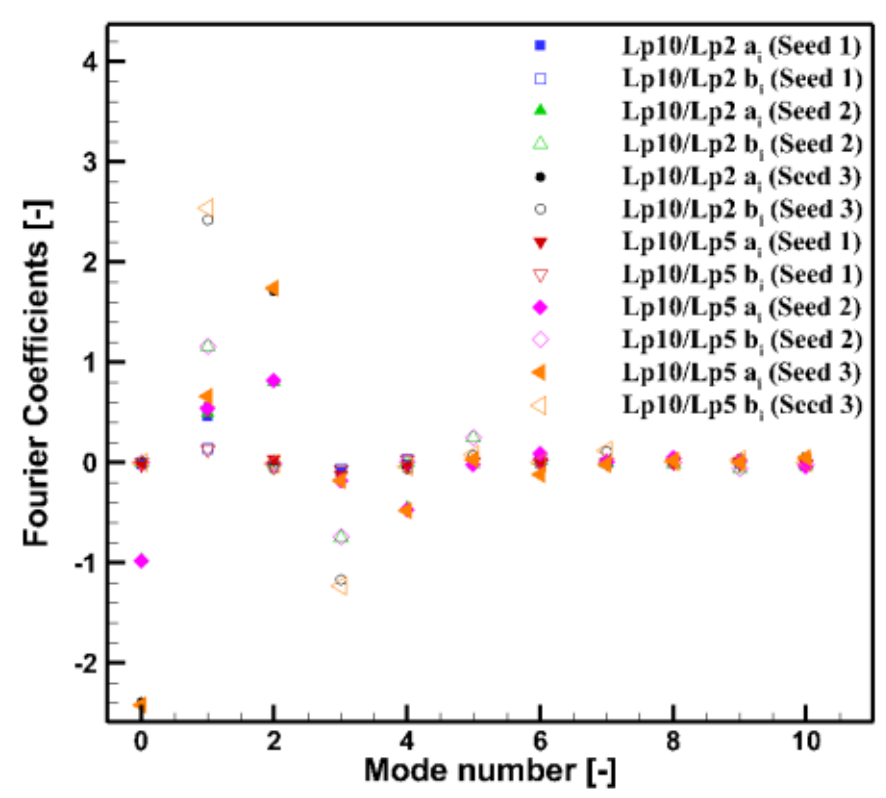

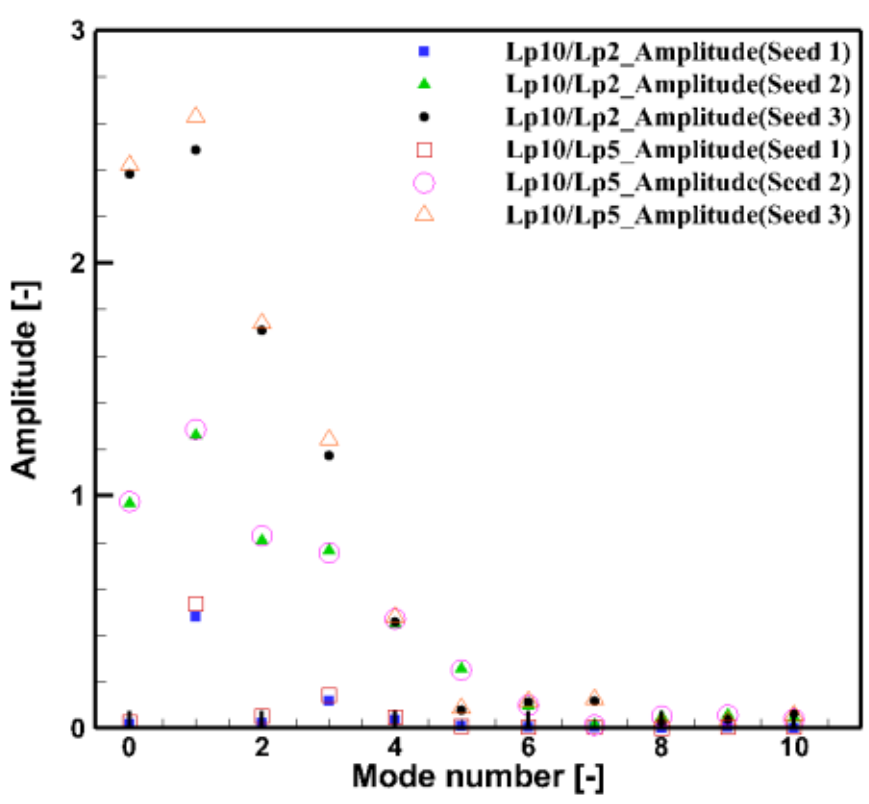

Figure 11. The Fourier coefficients (left image) and amplitude (right image) obtained from normalized stress plotted against mode number for two double networks with three seed numbers at 100% maximum strain amplitude

## Conclusion

The simulations have been done for three single network and two relevant double networks. The stress-strain relation showed little difference for the single network when the seed number changes at 5% maximum strain amplitude, while there were some more differences among the results for nonlinear regime, i.e. 100% maximum strain amplitude. For double network, the double peaks were not observed for some cases where the fibre alignments and seed number have effects on the results, not just hydrogen bonds (the crosslinkers) between networks, previously reported in the literature.
The Fourier coefficients (an identification of shear moduli) and resultant amplitude were evaluated for both single and double network. It was found that only the first mode must be considered in linear shear regime (5% maximum strain amplitude) for a single network. However, from zero until fifth modes must be considered for the exact single network in nonlinear regime, i.e. 100 % maximum strain amplitude. For double networks, in addition to the first mode, the zero mode must also be considered when the maximum strain amplitude is 5%. In nonlinear regime, from zero to fifth mode are required to capture the waveform in Fourier analysis. These results can help better understand the mechanical behaviour of these structures under shear flow, used in some applications such as tissue engineering and wound healing products.